\documentclass{iopjournal}

\usepackage{natbib}
\usepackage{amsmath}
\usepackage{mathtools}

\begin{document}

\articletype{Letter} 
\title{Climate change impacts on net load under technological uncertainty in European power systems}

\author{Luna Bloin-Wibe$^{1}$\orcid{0000-0002-5104-9168}, Erich Fischer$^1$, Leonard Göke$^{2}$, Reto Knutti$^1$, Francesco De Marco$^2$ and Jan Wohland$^{3}$}

\affil{$^1$Institute for Atmospheric and Climate Science, ETH Zurich, Zurich, Switzerland}

\affil{$^2$Reliability and Risk Engineering, ETH Zurich, Zurich, Switzerland}

\affil{$^3$Department of Technology Systems, University of Oslo, Kjeller, Norway}

\email{luna.bloin-wibe@env.ethz.ch}


\begin{abstract}

Renewable energy sources play a major role in future net-zero energy systems. However, achieving energy system resilience remains challenging, since renewables strongly depend on weather fluctuations, and future energy systems are subject to major design uncertainty. Existing literature mostly treats these types of uncertainty separately. Therefore, the assessment and comparison of uncertainties surrounding climate change and energy system design, and particularly their interactions, is insufficiently understood.

To close this gap, we evaluate net load to assess energy system stress without relying on perfect foresight, while maintaining the temporal and spatial correlations of the climate system. Net load is calculated from hourly historical and future climate model data translated to energy variables. To scope the extent of plausible energy systems, we consider eight different design scenarios inspired by the European Ten-Year Network Development Plan (TYNDP) and different levels of transmission expansion.

We find that climate change impacts on net load are highly sensitive to the energy system design, implying that energy systems can be designed so that they are either hindered or helped by climate change. Furthermore, within an energy system scenario, climate change can change the frequency and seasonality of high net load events and their technological and meteorological composition. Wind-dominated systems with currently electrified heating levels, for instance, feature a 30\% increase of high net load events under climate change, mostly in summer and fall, while fully electrified net zero systems are impacted by high net load events in winter and spring, which decrease by 50\% with climate change. Our work thus calls for a wider perspective on energy-climate stress that captures the non-linear interactions of climate change and energy system design uncertainty, thereby overcoming the current focus on cold "Dunkelflauten".

\end{abstract}

\section*{Introduction}\label{intro}
To limit global warming to 1.5°C above pre-industrial conditions, it is necessary to transition to net-zero renewable energy systems \citep{de_coninck_strengthening_2018}. However, the most important renewable sources, wind and solar, as well as the electrification of heating and cooling, increase the impact of meteorological conditions on electric demand and renewable generation \citep[e.g.,][]{staffell_increasing_2018,collins_impacts_2018}. Therefore, the resilience of renewable energy systems is subject to the dual uncertainty of the energy system configuration and the effects of climate change. 

The existing literature mostly treats climate change and energy system design impacts separately. Several studies use climate models to assess changes in climate and variability of the generation potential of solar PV, wind, hydropower and demand \citep{cronin_climate_2018,kapica_potential_2024, plaga_methods_2023, tobin_vulnerabilities_2018, yalew_impacts_2020}. However, these studies do not fully explore the role of the energy system, which includes not only the generation potential of each technology, but also its installed capacity, spatial distribution, and the importance of each technology compared to the others. 

On the other hand, optimized renewable energy systems that are resilient to the variability of weather conditions, have also been extensively studied \citep{ hilbers_importance_2019, li_stochastic_2016, de_marco_climate-resilient_2025,panteli_influence_2015, perera_quantifying_2020, plaga_methods_2023}. However, due to computational constraints, energy system optimization usually rely on data reduction techniques when handling climate and weather data: this can take the form of optimizing single years separately, using only historical years rather than taking into account the climate change signal of the recent and coming years, or synthetically generating a few typical or extreme weather years \citep{plaga_methods_2023, rouges_link_2025}. Methods like importance subsampling, based on a longer time series, are also employed, which do preserve the spatial, but not the temporal consistency of the original climate simulations \citep{hilbers_importance_2019}. Furthermore, the field of Modeling to Generate Alternatives (MGA), applied to energy systems, has shown the importance of considering a larger system design space, rather than a single cost-optimal energy system \citep{decarolis_modelling_2016, pickering_diversity_2022}.

Additionally, to optimize the energy system in terms of capacity investment and optimal operation, these energy system models usually rely on perfect information about weather conditions far into the future, so-called perfect foresight. Approaches with more limited foresight exist, but the time frame for which information is available is still at least one year \citep{heuberger_impact_2018, mannhardt_understanding_2024}, with a few notable exceptions \citep{goke_liquid_2025}. The realism and interpretability of these models are thus limited because real-world deterministic weather predictability is restricted to a few weeks \citep{krishnamurthy_predictability_2019}. This could, for example, cause unrealistic long-term storage operation as noted in \citet{dowling_role_2020}, where different foresight horizons lead to different critical time steps and filling levels despite identical weather evolution.

Efforts have been made to overcome the separation between energy and climate modeling \citep{craig_overcoming_2022}. For example, studies on multi-year climate model data, translated into energy-relevant variables, often incorporate ways of inter-comparing technologies \citep{bloomfield_meteorological_2020, van_der_most_temporally_2024, van_der_wiel_meteorological_2019} and considering different spatial distributions \citep{kittel_coping_2025}. Some studies have combined energy system optimization models with such multi-year climate model data \citep{grochowicz_using_2024, schlott_impact_2018, simoes_climate_2021, wohland_climate2energy_2025}, although limitations like
perfect foresight and single energy system design remain. The direct assessment of uncertainties surrounding both climate change and the energy system configuration is -- to the best of our knowledge -- studied only by \citet{bloomfield_quantifying_2021}, a pioneering work where the technological scope (a wind-solar-demand system with a copperplate assumption) and temporal resolution (seasonal to annual) place limits on possible conclusions. We thus lack a comprehensive understanding of the interplay of energy system design uncertainty and climate change.

In this paper, we evaluate net load, or the remaining electric demand after renewable generation, as a simple metric to assess whether the energy system is under stress. In doing so, we answer the question of what the combined effects of the energy system configuration and climate change on the European generation and demand are, and in what ways they are intertwined. 


\section*{Methods}\label{methods}

\subsection*{Converting climate model data into energy-relevant variables}
Climate models provide simulations of physically consistent weather conditions and their change in response to increasing greenhouse gas concentrations, changes in aerosol concentrations and land-use change, as well as variations in solar and volcanic forcing. In this paper, we define two climate periods that will be directly compared: a historical climate period (1995-2015) and an end-of-century climate period (2080-2100; SSP3-7.0), both generated from the state-of-the-art fully coupled Community Earth System Model 2.1.2 \citep{danabasoglu_community_2020}. For each climate period, we get a better sample of climate variability by generating three parallel simulations, or ensemble members, from the same climate model that only differ by their initial conditions. Thus, our experiment yields a total of 60 years per climate period.

To obtain tailored energy-relevant generation and demand variables stemming from this climate model data, we employ Climate2Energy \citep{wohland_climate2energy_2025}. This modular pipeline performs bias correction, before calculating country-level datasets of on- and offshore wind power, solar PV and electric heating and cooling demand (using established open-source tools), as well as run-of-river and reservoir hydropower through a hydropower model based on internally routed river discharge.

\begin{figure}[h]
 \centering
        \includegraphics[width=0.9\textwidth]{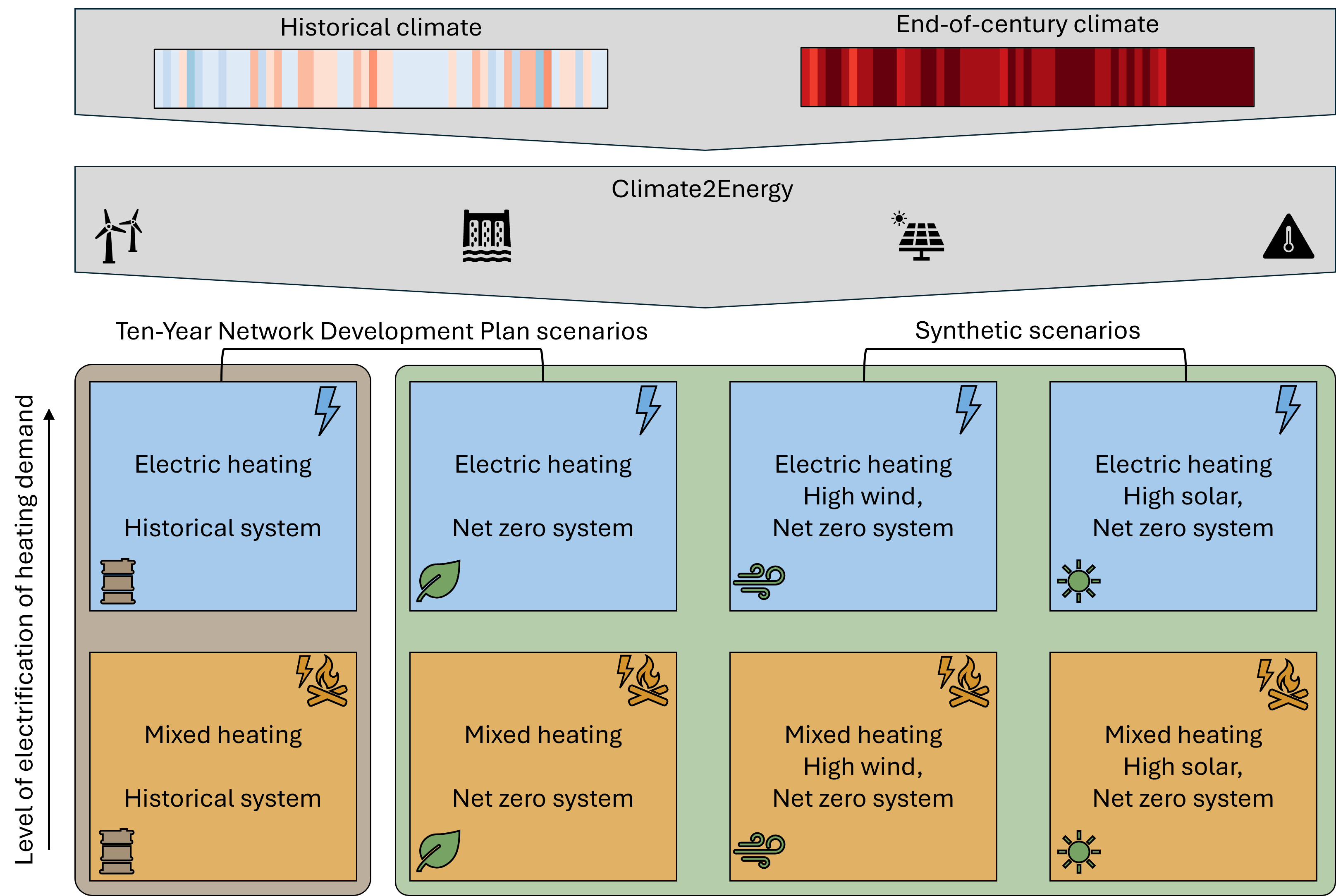}
 \caption{\textbf{An illustration of the climate data conversion process through Climate2Energy and the eight different energy system scenarios used in this study.} The first row (blue boxes) represents the scenarios that have fully electrified heating demand, while the second row (orange boxes) represents the scenarios that have heating demand corresponding to the current share of electrification in each country. The brown background shading in the first column represents the Ten-Year Network Development Plan (TYNDP) historical system setup (highlighted by a barrel icon), while the green background shading in the 2nd to last column shows the net-zero system setups: the official TYNDP net-zero system setup, highlighted by a leaf icon; the synthetic high wind setup, highlighted by a wind icon, and the synthetic high solar setup, highlighted by a sun icon. The historical and end-of-century climate period is symbolized by temperature warming stripes calculated from the data set.}
\label{fig:scenario_configs}
\end{figure}

\subsection*{Matrix of energy scenarios}

To compare different energy systems in a broad but systematic way, while keeping the number of scenarios reasonably low, we evaluate the space of plausible system designs schematically. By combining different setups of heating electrification and wind-to-solar generation share, we obtain a 2 by 4 scenario matrix (see Figure \ref{fig:scenario_configs}). This way, we can compare historical systems to future (net-zero) systems, allowing us to analyze challenges and opportunities. We can also assess different net-zero setups to account for design uncertainty of future energy systems. The two electrification setups additionally provide boundaries of plausibility for the importance of electrified heating demand in the energy system. 

We use all energy system scenarios with data from both historical and end-of-century climate periods, enabling a direct assessment of climate change impacts on various plausible energy system scenarios.

\subsection*{Net load calculation}  \label{sec:nl}

The net load $N$ of a given energy system scenario at a time $t$ can be calculated as $N(t) = D(t)-G(t)$, where $D$ is the total electricity demand and $G$ is the total renewable generation. Positive net load means that electricity demand is higher than renewable generation, and therefore indicates stress in energy systems dominated by renewables, because non-renewable sources or storage must contribute to meet demand. By design, therefore, net load does not take into account non-renewable generation sources nor storage technologies, with the exception of hydropower, due to its simultaneous use as a generation and storage source (see below).

The simplicity of the net load metric thus allows for a robust assessment of energy system stress, where both climate change and the configuration of the energy system can be varied and assessed. Importantly, it also maintains the temporal and spatial correlations of the climate system that could be lost using data reduction techniques, and does not rely on unrealistic perfect foresight assumptions generally made in energy system optimization models.


\subsubsection*{Generation}

Our inputs from Climate2Energy \citep{wohland_climate2energy_2025} provide the fraction of potential renewable power output, or capacity factors (CF), for wind and solar energy. Therefore, we need to make assumptions on the installed capacity of each technology to get the absolute generation necessary to compute net load. To account for the projected expansion of on- and offshore wind and PV, we use values for historical and projected future capacities from the TYNDP 2024 Scenarios Report \citep{entso-e_tyndp_2024}, for each country separately. The chosen future system setup, corresponding to the Global Ambition scenario in the TYNDP report, is designed to achieve carbon neutrality by 2050. 


 Additionally, we artificially vary the relative role of wind and solar in future energy systems, reflecting the large cost and social support uncertainty surrounding the actual deployment of these key technologies. We define two additional variants of the net-zero system setups by changing the relative role of PV and on- and offshore wind capacity in a way that keeps total mean annual generation $G$ unchanged relative to the original net-zero system setup, $\widetilde{G}$. 

\begin{equation}
    G = \sum_{i,c} G_{i,c} = \sum_{i,c} IC_{{i,c}} \cdot \overline{CF}_{{i,c}} \coloneq \widetilde{G}, \qquad i \in[\mathrm{PV},\mathrm{W_{on}},\mathrm{W_{off}}],
\end{equation}

where $IC_{i,c}$ is the installed capacity and $\overline{CF}_{{i,c}}$ is the average CF for technology $i$ and country $c$, covering all European countries. 

For the other generation variables, run-of-river and reservoir hydropower, no substantial capacity expansion is projected, and the direct Climate2Energy output in GWh is used invariably for all setups. 

\subsubsection*{Demand}

We use demand for electric heating and cooling from Climate2Energy \citep{wohland_climate2energy_2025}, which internally leverages the demand.ninja model \citep{staffell_global_2023}. Climate2Energy provides two different heating demand scenarios: the first is based on the currently electrified heating share and the second assumes full electrification of heating demand. Despite the push towards electrification of the heating sector, a fully electrified heating sector is an extreme scenario. We therefore use both electrification setups to probe the space of future electrical heating demand. 

To calculate the weather-insensitive demand, a similar approach is taken to that of the generation CFs: normalized time series are multiplied by absolute values (in GWh) for each country according to the TYNDP 2024 Scenarios Report, for a historical and a net-zero future setup. To do this, we follow the approach of \citet{goke_how_2023}, which is detailed in the supplementary material. 

\subsection*{Hydropower dispatch model} \label{sec:hydro}

Reservoir hydropower typically serves a dual purpose within current European energy systems, as it can be used both to generate and store electricity. In order to include the storage capabilities of reservoir hydropower, without resorting to the perfect foresight assumption necessary to run a full optimization, we create a heuristic dispatch model. 

When net load is negative, all inflow is added to the reservoir storage because demand is already met by variable renewables. When net load is positive, indicating a gap in supply, the stored hydropower is used to lower net load. We constrain generation from hydropower based on installed turbine capacities and reservoir filling level relative to its average seasonal evolution. Our approach reflects the concept of risk-aversion to avoid depleted reservoirs, by limiting usable hydropower capacity to 50\% when the storage filling level is within one standard deviation of the usual values on a given calendar day, further reducing to 10\% when the filling level drops below this range. We provide an in-depth assessment of the dispatch model in the supplementary material, showcasing that the dispatch model simulates hydropower storage effects to a satisfying degree, decreasing peaks in net load also during periods of low reservoir inflow.

\subsection*{Transmission effects between countries} \label{sec:transm}

An assessment of net load over Europe requires removing the per-country dimensionality. One of the simplest way to do this would be to sum the net load over all countries, thus assuming that all countries could transfer any excess generation to another instantaneously (a so-called copperplate assumption). This simple framework has been used in other studies \citep{ gotske_cost_2023, kittel_measuring_2024, priesmann_are_2019}, since it establishes an upper bound of the transmission benefits and does not introduce assumptions on the transfer capacities between countries or the timing of transfers. However, the copperplate assumption is inherently unrealistic, in particular since periods of high net load can be exacerbated by limited transmission between countries \cite{grochowicz_using_2024, kittel_coping_2025}. 

Therefore, we develop a simple transmission model based on \citet{wohland_more_2017}: energy can be transmitted between countries with the objective of minimizing European total net load, constrained by the technical limits of the existing Net Transfer Capacities (NTC) \citep{entso-e_entso-e_2019}.  Underlying equations and an assessment of the simplified realistic transmission model can be found in the supplementary material. We report that the model efficiently transfers excess generation between countries, leading to an exacerbation of certain periods of net load compared to the copperplate assumption. 

However, since transmission occurs when net load is above zero, without considering non-renewable generation and non-electric demand, it could also lead to unrealistic effects, in particular for the historical system, where non-renewables still play a relevant role in the total system. We therefore show results for both frameworks: the simplified realistic transmission will be used when studying high net load events (i.e., when it is important to differentiate between events that all have high net load), and the copperplate assumption when evaluating  net load distributions (i.e., when it is important to differentiate between all net load events, many of which could be below zero).

\section*{Results}

\begin{figure}[h]
 \centering
        \includegraphics[width=\textwidth]{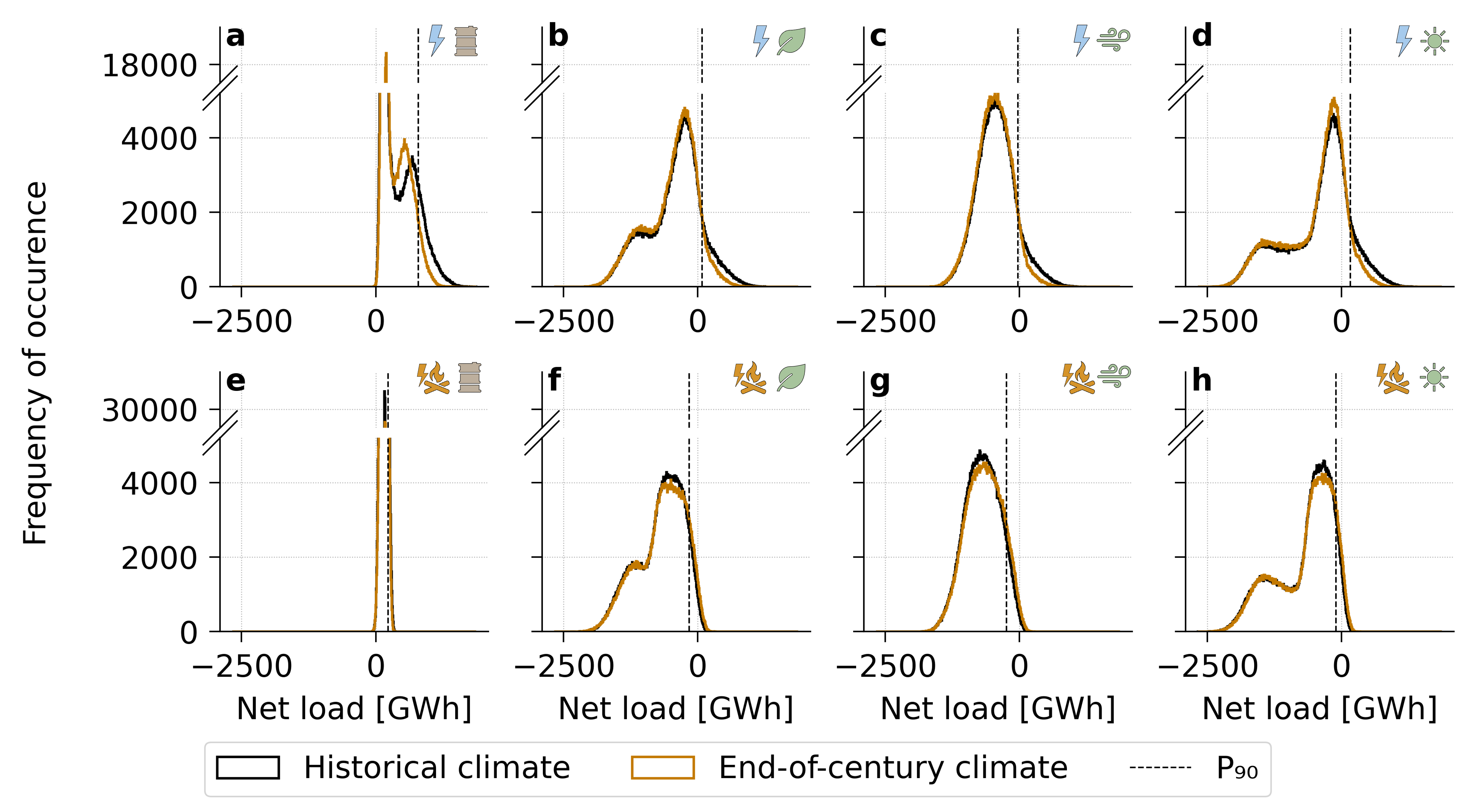}
 \caption{\textbf{Distribution of mean European net load over all available time steps  in different energy scenarios for the copperplate assumption.} Results for historical climate conditions are shown in black and results using end-of-century climate are shown in orange. Subpanels are organized as in Figure \ref{fig:scenario_configs}, icons in the upper-right corner representing energy scenario choice. They correspond to the (\textbf{a-d}) electric heating and (\textbf{e-h}) mixed heating setups, and the (\textbf{a,e}) historical system, (\textbf{b,f}) net-zero TYNDP system, (\textbf{c,g}) net-zero, high wind system and (\textbf{d,h}) net-zero, high solar system setups. The black dotted line in each panel shows the net load 90$^{\mathrm{th}}$ percentile across the two climate periods.
 }
\label{fig:nl_distrib}
\end{figure}

\subsection*{The vulnerability of energy systems to climate change is design-specific}

Figure \ref{fig:nl_distrib} shows that the distribution of net load is more affected by changes in the energy system scenario (different panels) than by climate change (black and yellow lines). 

The largest difference lies between the historical and the three future net-zero system setups: in the historical setup (Panels \textbf{a,e}), most values are above zero, since the share of renewables in these systems is low. In the latter setups  (Panels \textbf{b-d,f-h}), most values are below zero, but the increased reliance on renewables also widens the distribution of net load, since they are more susceptible to short-term fluctuations. 

Another substantial difference lies in the heating electrification setups. The net load in all electric heating demand setups  (Panels \textbf{a-d}) exhibit a longer positive tail than in the mixed heating demand setups (Panels \textbf{e-h}), due the importance of heating demand in the total net load.
This tail is also where the climate change effects are the most visible across energy scenarios: electric heating demand setups show substantially fewer high net load events above the 90$^{\mathrm{th}}$ percentile by end-of-century, while the mixed heating demand show a slight increase. We thus see that an energy system scenario can be chosen so that the system is either hindered or helped by climate change. 

\subsection*{Climate change alters the seasonality and frequency of high net load events}

\begin{figure}[h]
 \centering
        \includegraphics[width=\textwidth]{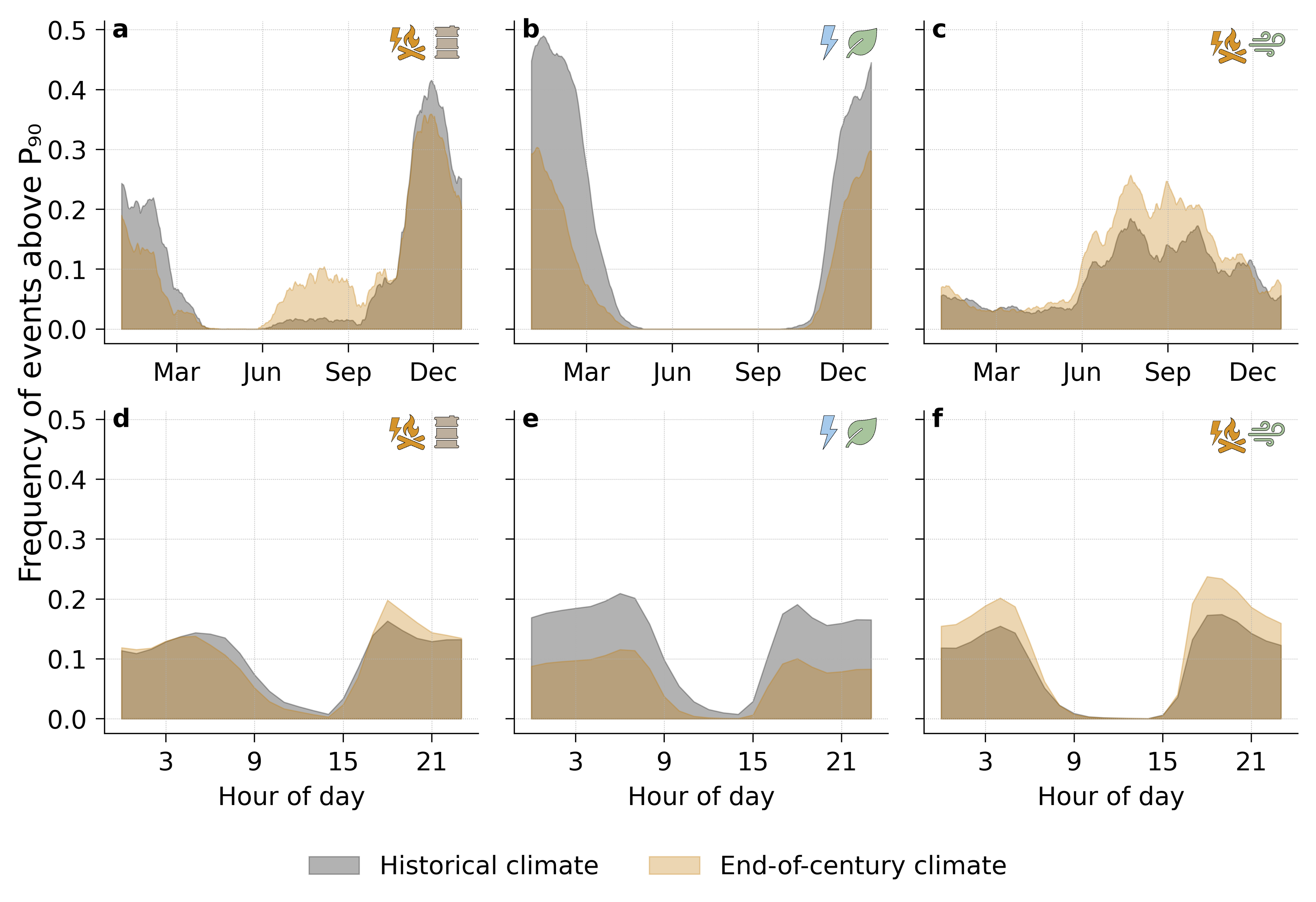}
 \caption{\textbf{(a-c) Seasonal and (d-f) diurnal cycle of high net load events (above the 90$^{\mathrm{th}}$ percentile, $\mathrm{P}_{90}$ ) for three different scenarios, with simplified realistic transmission.} Results for historical climate conditions are shown in black and results using end-of-century climate are shown in orange. The different scenarios are represented by icons in the upper-right corner and correspond to (\textbf{a,d}) "historical system; mixed heating", (\textbf{b,e}) "net-zero system; electric heating" and (\textbf{c,f}) "net-zero system, high wind; mixed heating".
 }
\label{fig:daily_season_peaks}
\end{figure}

While smaller than the impact of the energy scenarios themselves, we find discernible climate change impacts in all scenarios. To improve our understanding of the effects of climate change in the tails of the net load distribution, we hereby focus on high net load events. They are selected by calculating the 90$^{\mathrm{th}}$ percentile jointly for both the historical and end-of-century climate periods, so that the frequency of high net load events can be different between climate periods. However, due to the variation in net load between scenarios, high net load events are selected for each scenario separately. 

Figure \ref{fig:daily_season_peaks} shows the seasonal and diurnal distribution of high net load events for three representative scenarios (see the supplementary material for an overview of all scenarios). In all three scenarios, we find climate change effects in the seasonal distribution of high net load events (Figure \ref{fig:daily_season_peaks}\textbf{a-c}). However, the type of change strongly depends on the energy system scenario. 

Firstly, in the "historical system; mixed heating" scenario (Panel \textbf{a}), we note a temporal shift in high net load events towards summer: while they almost exclusively occur in winter and fall (94\%) in the historical period, the winter occurrence drops to (80 \%) by end of century, in favor of an increase in summer high net load events (17\%). However, the total number of high net load events stays stable between climate periods.

In the two other scenarios (Panels \textbf{b} and \textbf{c}), seasonality does not qualitatively change between historical and future climate. Instead, the overall frequency of events changes: for the "net-zero system; electric heating" scenario, high net load events decrease from the historical period to half as many in the end-of-century period, while in the "net-zero system, high wind; mixed heating" scenario, high net load events increase by 1.3 times from the historical to the end-of-century period. 

In contrast to the seasonal shift seen in Panel \textbf{a}, there is no temporal shift in the diurnal cycle between the historical and future climate (Figure \ref{fig:daily_season_peaks}\textbf{d-f}). Indeed, all scenarios and climate periods show the same overall picture: there is substantial diurnal variability, with minima between 9am and 3pm, when solar radiation is at its peak. This effect is more pronounced in Panels \textbf{e} and \textbf{f}, where the energy system scenario has a higher adoption of solar PV. 

\subsection*{Climate change alters the technology composition of high net load events}

\begin{figure}[h]
 \centering
        \includegraphics[width=\textwidth]{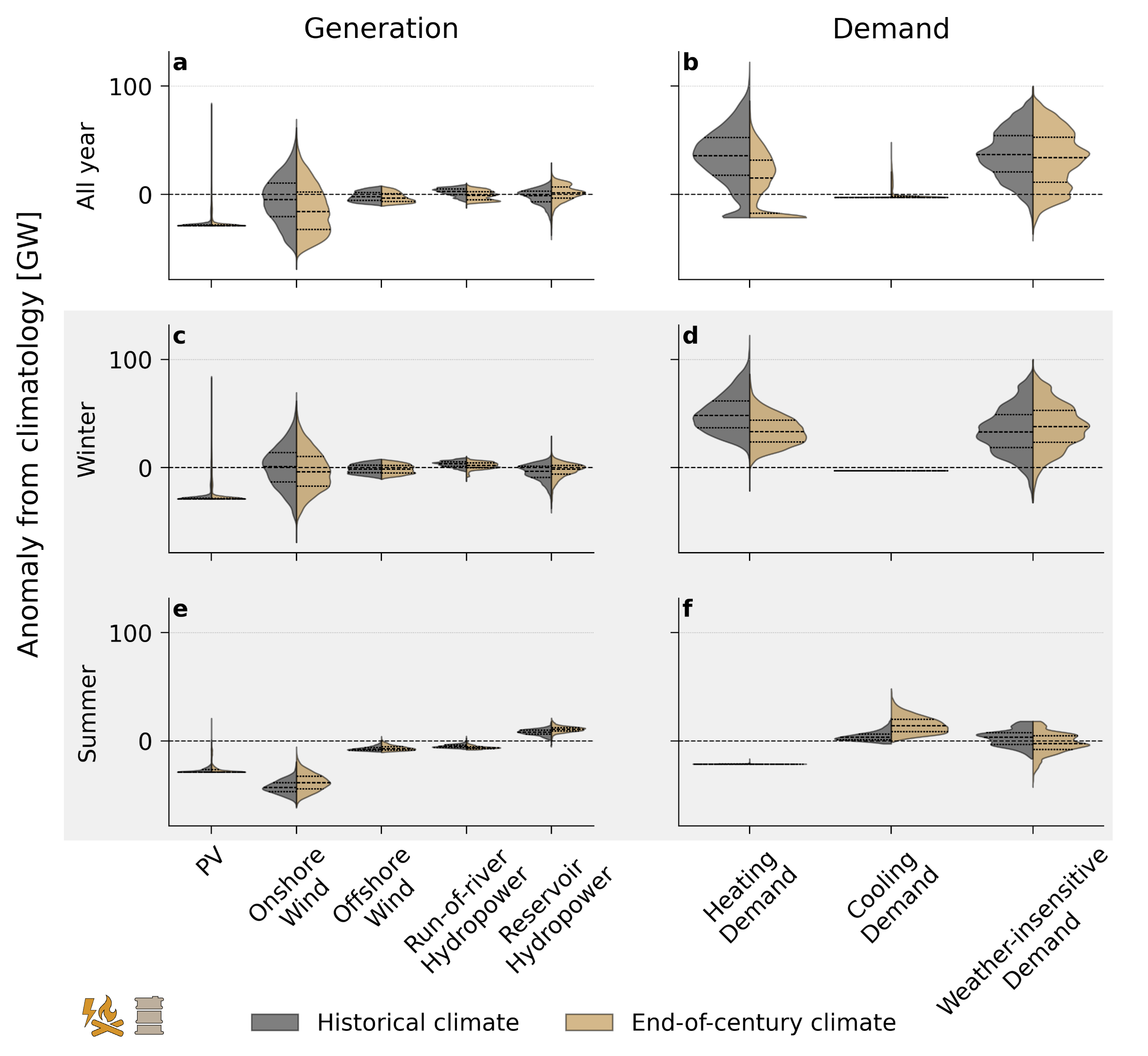}
 \caption{\textbf{Violin plots of (a,c,e) generation and (b,d,f) demand anomaly for high net load events (above $\mathrm{P}_{90}$), for (a-b) all seasons, (c-d) winter and (e-f) summer.} Results for historical climate conditions are shown in black and results using end-of-century climate are shown in orange. Anomalies are calculated over all available time steps in both climate periods. The scenario corresponds to "Mixed heating; historical system" (represented by the bottom left icons), using simplified realistic transmission. A negative generation and positive demand technology anomaly means that it contributed to the stress of the event.
 }
\label{fig:violins}
\end{figure}

To improve process understanding, we now decompose the contribution of different technologies to high net load events. We focus on the "net-zero system; electric heating" scenario. For clarity, we show results separately for generation (Figure \ref{fig:violins}\textbf{a,c,e}) and demand (Figure \ref{fig:violins}\textbf{b,d,f}), and we report anomalies. A negative \textit{generation} anomaly implies that there is less generation than usual, thus contributing to the stress of the event. On the other hand, a negative \textit{demand} anomaly means that demand technology was needed less than usual, thus alleviating the stress of the event. 

For all seasons combined (Panels \textbf{a,b}), we find that most high net load events present PV generation deficits and high heating and weather-insensitive demand, while the other technologies have distributions spreading across both positive and negative anomalies. Climate change somewhat alters this picture: compared to the historical climate period, heating demand is much less important for the high net load events in the end-of-century period as a result of the warming climate, while cooling demand, on- and offshore wind and run-of-river deficits become more important. 

However, when looking at high net load specifically in winter (Panels \textbf{c,d}) and summer (Panels \textbf{e,f}), two distinct types of high net load events emerge. In winter, we observe that weather-insensitive and heating demand, as well as negative PV anomalies are the strongest factors contributing to high net load events. Onshore wind deficits, and to a lesser extent offshore wind and run-of-river and reservoir hydropower deficits, can also be contributing factors, but this is not always the case. On the other hand, we find that high net load events in summer are driven by wind and PV deficits, as well as higher than average cooling demand.

We can thus disentangle the seasonal climate change effects from those that affect high net load events regardless of the season. For example, the climate change-induced increase in wind generation deficits in high net load events actually masks two opposing trends: in winter, wind deficit becomes a stronger contributor to high net load events, while the opposite is true in summer, both changes being statistically significant (calculated with a 2-sample Kolmogorov-Smirnov test). Furthermore, the apparent bimodality of heating and demand anomalies over all seasons can be explained by the different seasonal behavior observed in winter and summer.

This description holds true only for the "mixed heating; historical system" scenario, highlighting the need for design-aware climate change impact assessments. The supplementary material shows the same violin plots for "mixed heating; high wind, net-zero system" and "electric heating; net-zero system”, for which the generation and demand composition shows a decrease in heating demand anomaly and an increase in on- and offshore wind deficit with climate change.

\begin{figure}[h]
 \centering
        \includegraphics[width=\textwidth]{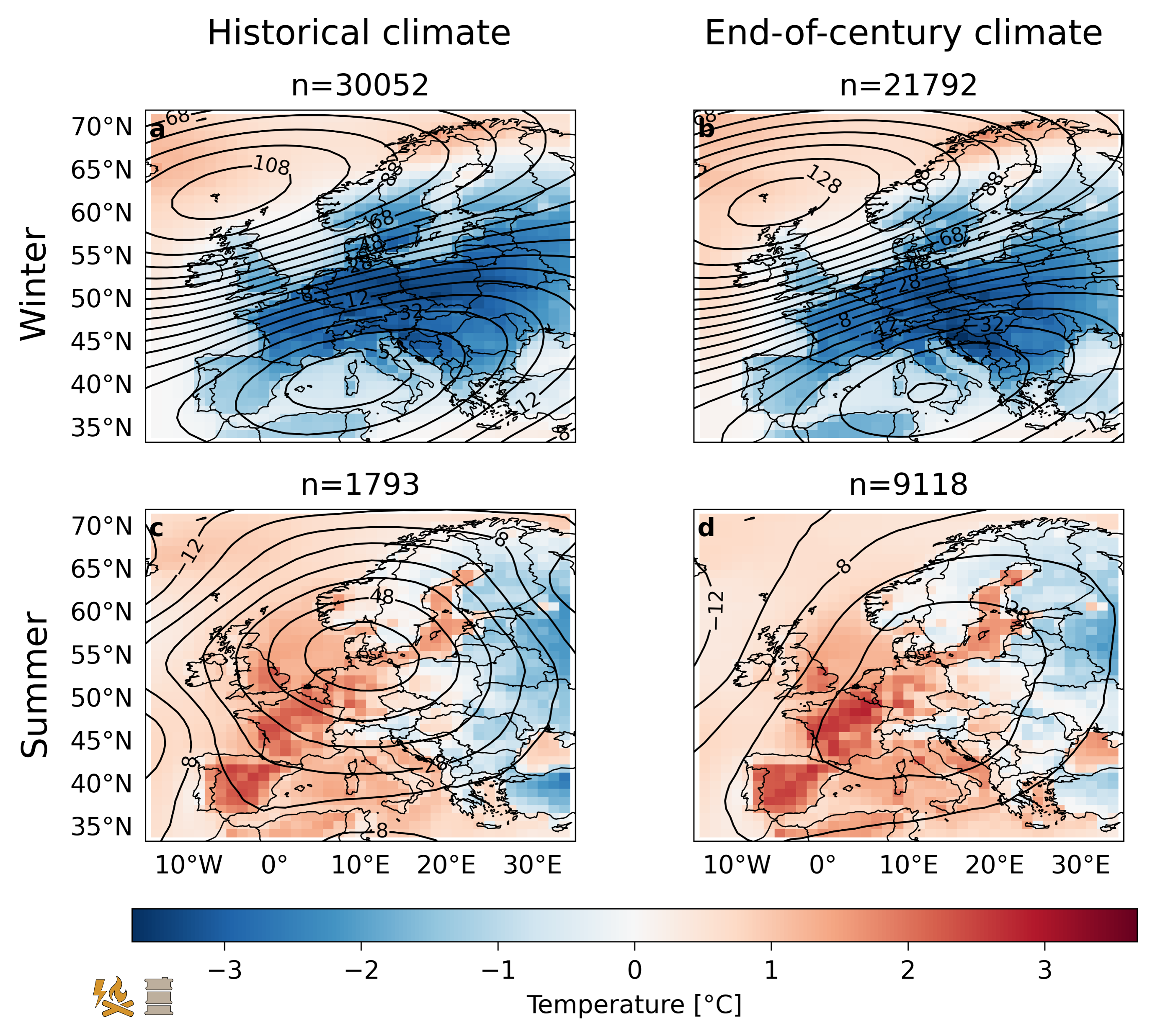}
 \caption{\textbf{Composites of seasonal anomalies of temperature and 500 hPa geopotential height contour lines for high net load events (above $\mathrm{P}_{90}$) in (a-b) winter and (c-d) summer, for the (a,c) historical and (b,d) end-of-century climate periods.} Anomalies are calculated from all available time steps in both climate periods, for each season separately. The scenario corresponds to "Mixed heating; historical system" (represented by the bottom left icons), using simplified realistic transmission. n refers to the number of events included in each map.
 }
\label{fig:weather}
\end{figure}

\subsection*{Weather patterns for high net load events are moderately altered by climate change}

Figure \ref{fig:weather} gives insights into the weather patterns present during high net load events in the scenario "Mixed heating; historical system". It shows temperature and geopotential height anomalies in winter (Panels \textbf{a,b}) and summer (Panels \textbf{c,d}) relative to the respective seasonal mean, for the historical (Panels \textbf{a,c}) and end-of-century (Panels \textbf{b,d}) climate periods. We compute anomalies with respect to each climate period's seasonal climatology. 

We find that high net load events in winter experience similar weather conditions in both climate periods; indeed, the Pearson pattern correlation between future and historical anomaly patterns is above 0.98 for both variables. In both climate periods, we find the European continent colder than usual, in particular in central Europe. This cold anomaly is associated with a positive geopotential height anomaly over the Northern Atlantic, and a negative geopotential height anomaly over the Mediterranean, leading to less westerly flow or even easterly advection. The weather conditions corresponds to a so-called “Kalte Dunkelflaute”, when high pressure systems co-occur with anomalously low winds, and cold anomalies cause an increase in heating demand. Combined with the lack of PV generation in winter, this leads to high values of net load.

High net load events in summer show a different weather picture. Here, a large-scale high pressure anomaly covers central Europe, and we find abnormally high temperatures in western and southern Europe. This weather situation does present larger differences between climate periods than in winter. Temperature anomalies increase slightly in southwestern Europe by the end of the century, and the geopotential height anomaly weakens. This leads to a lower pattern correlation than in winter, with 0.94 for temperature and 0.75 for geopotential height anomalies.

\section*{Discussion and conclusions}


We have shown that both energy system design and climate change have discernible impacts on net load, and that they are non-trivially linked: in particular, we report that climate change effects are highly sensitive to the energy system design, with some systems showing less high net load events by the end of the century, and others showing more. 

While its effects are relatively small for a given energy scenario, climate change can still alter the frequency, seasonality and composition of high net load events. Furthermore, energy system scenarios and climate change differ in terms of human agency: the former is a result of socio-political design decisions in one region, while the latter is governed by complex global human-Earth System interactions. Since any chosen energy system needs to be able to cope with future climate, these climate change effects are still crucial to consider.


We find that for a given energy scenario, there are specific meteorological and technological patterns that lead to high net load events, which are differently altered by climate change. In particular, in the "Mixed heating; historical system" scenario, we observe two distinct types of high net load events in winter and summer. In winter, we identify a "Kalte Dunkelflaute": an anticyclonic anomaly north of the British Isles reduces the westerly winds and warm air advection, leading to anomalously cold temperatures across most of the European continent (Figure \ref{fig:weather}\textbf{a-b}). This result is consistent with the high importance of heating demand and wind deficits found in Figure \ref{fig:violins}\textbf{c,d}. With climate change, these processes do not fundamentally change; rather, high net load events will rely less on abnormally high heating demand (due to the climate change signal), and more on wind deficits, as reflected in the increased geopotential height anomaly over the Northern Atlantic in the end-of-century period, as well as the technology changes in Figure \ref{fig:violins}\textbf{c,d}.

In summer, we find that high net load events occur under different weather conditions (Figure \ref{fig:weather}\textbf{c-d}). Firstly, the large-scale high pressure anomaly over Central Europe corroborates the high wind deficits found in Figure \ref{fig:violins}\textbf{e}, since high pressure systems are generally associated with low wind speeds at their center. Secondly, the abnormally high temperatures in western and southern Europe, where air-conditioning is wide-spread, align with the higher than average cooling demand found in Figure \ref{fig:violins}\textbf{f}. Climate change further amplifies summer high net load events: despite a weaker geopotential height anomaly, climate change induces higher temperatures, and thus more pronounced cooling demand, which becomes a dominant contribution to these events, whereas wind deficits become somewhat less important. When considering a broad range of plausible energy systems, we conclude that high net load events are not limited to the well-known challenges of "Kalte Dunkelflauten". Indeed, we find high net load events also in summer, which present different technological and meteorological features and an increase in frequency with climate change. 


No storage technologies, apart from hydropower reservoirs, were accounted for in this study. By looking at when and where net load is high, we can therefore identify storage needs to reduce the number of high net load events. When comparing daily and seasonal time scales (Figure \ref{fig:daily_season_peaks}), for example, we find that diurnal fluctuations consistently increase in all net zero system setups compared to the historical energy system setups. This means that the challenge of short term storage must be addressed regardless of the chosen future energy system design. On the other hand, while all scenarios show seasonal fluctuations, their magnitude and timing vary substantially. It is therefore harder to determine how seasonal storage should be managed optimally given the energy scenario uncertainty.


We conclude that assessing the effect of climate change or system design on energy systems cannot be performed in isolation, because doing so creates the risk of missing important intertwined effects. Future studies should therefore incorporate both design and climate change uncertainty and their interactions when addressing energy system resilience. Furthermore, they could expand on the insights found in this paper by refining methods and models. We share our code openly to enable such future work. For example, adding varying levels of cooling infrastructure adoption could help better understand the uncertainties around summer events, and incorporating data from multiple climate models could quantify the uncertainty around climate system response to increased greenhouse gases. Other promising avenues include further analyzing the role of energy storage during high net load events across different energy system scenarios, and exploring storylines in the context of high net load events. This would permit a qualitative understanding of key aspects of energy system stressors, like drivers, predictability, duration and possible cascading effects. 


\ack{LBW, LG, RK, FM and JW are part of the SPEED2ZERO, a Joint Initiative co-financed by the ETH Board. We thank the National Center for Atmospheric Research (NCAR) for developing the Community Earth System Model, and, since all analysis was carried out in Python, its contributors, as well as those who contributed to Python packages, in particular xarray, numpy and scipy.} 

\roles{LBW: conceptualization, data curation, formal analysis, methodology, software, visualization, writing (original draft preparation, review and editing); EF: conceptualization, supervision, writing (review and editing); LG: methodology, writing (review and editing); RK: conceptualization, funding acquisition, supervision, writing (review and editing); FM: data curation, writing (review and editing); JW: conceptualization, methodology,supervision, writing (review and editing).}

\data{All code (preprocessing, calculation and plots) is available at \url{https://github.com/luna-bloin/net_load_boosting}.}


\bibliography{references}

\section*{Supplementary material}

\subsection*{Synthetic net zero system setups}

To probe the space of possible energy scenarios, we define, in addition to the two official TYNDP scenarios, two synthetic net-zero system setups, which change the relative composition of installed PV and on- and offshore wind capacity ($W_{on}$, $W_{off}$ ), but constrain the total mean annual generation $G$ to that of the original net-zero system setup, $\widetilde{G}$.

Specifically, we define one high wind setup, which has two times as much wind generation as solar generation on average, and one high solar setup, which has two times as much solar generation as wind generation on average. Additionally, we set the fraction of on- and offshore wind capacity to be that of the TYNDP  in all three net-zero energy system setups. This leads to the following system of linear equations, where the unknowns, $IC_{\mathrm{W_{on}},c}$ and $IC_{\mathrm{W_{off}},c}$, can be determined for each country considered, separately:

\begin{align}
    G &=   \widetilde{G},\\
    x \cdot G_{\mathrm{PV},c} &= G_{\mathrm{W_{on}},c} + G_{\mathrm{W_{off}},c}, \\
    \frac{IC_{\mathrm{W_{on}},c}}{IC_{\mathrm{W_{off}},c}} &= \frac{\widetilde{IC}_{\mathrm{W_{on}},c}}{\widetilde{IC}_{\mathrm{W_{off}},c}},
\end{align}

where $IC_{i,c}$ is the installed capacity for technology $i$ and country
$c$ and $x$ is the scaling factor (either 0.5 or 2). 

\subsection*{Weather-insensitive demand calculation}\label{sec:wi_demand}

\begin{figure}[h]
    \includegraphics[width=\textwidth]{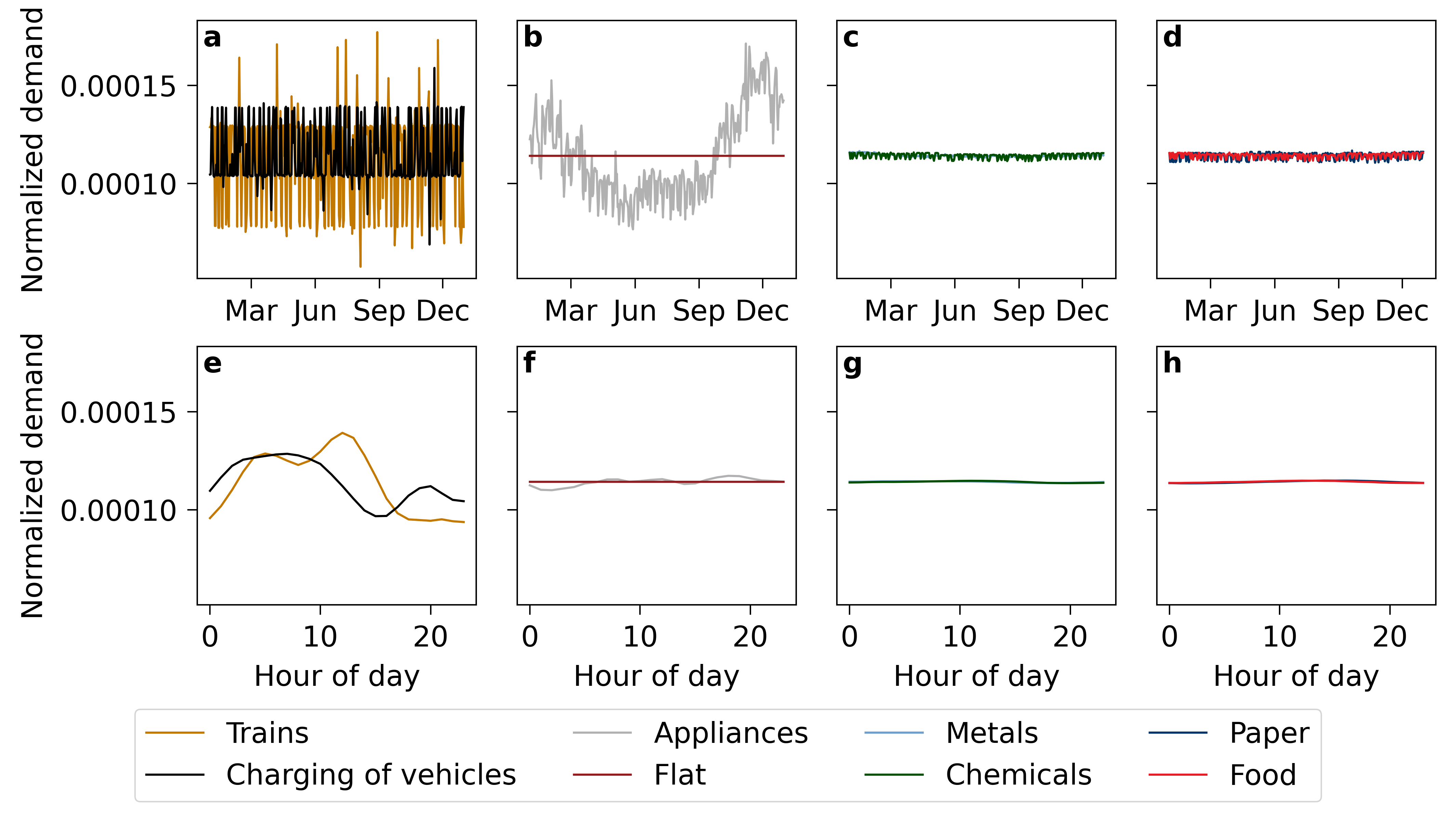}
    \caption{\textbf{The (\textbf{a-d}) yearly and (\textbf{e-h}) diurnal cycle of the normalized demand of all types of weather-insensitive demand types considered in this study.} (\textbf{a,e}) shows trains (freight and passenger) in yellow and charging of vehicles (busses, cars, trucks, vans) in black, (\textbf{b,f}) shows appliances in gray and the flat profile (used for data centers, refineries, national and international aviation and shipping, and others) in dark red, (\textbf{c,g}) shows metals (iron, steel, others) in light blue and chemicals in green, and (\textbf{d,h}) shows paper in light red and food (and fertilizers) in dark blue. }
    \label{fig:wi_demand}
\end{figure} 

To calculate the weather-insensitive demand, normalized time series are multiplied by absolute values for each country according to the TYNDP 2024 Scenarios Report, for a historical and a net-zero future setup. However, since the climate model by definition cannot provide  data for weather-insenstitive demand, we follow the approach of \citet{goke_how_2023}. Here, yearly profiles are taken from Plan4Res \citep{most_plan4res_2020} of electric demand for metals, chemicals, paper, food, charging vehicles and vehicles on the road, as well as exogenous demand for appliances. Other types of electric demand, like data centers and aviation, is assumed to have a flat yearly profile (for a detailed breakdown of the different types of demand, see Appendix figure \ref{fig:wi_demand}). These normalized profiles are calculated for the year 2015, but are assumed to apply for all time series considered in this study. Since the UK no longer forms part of the TYNDP 2024 Scenarios Report, values for this country were extracted from the TYNDP 2022 Scenarios Report \citep{entso-e_tyndp_2022}, in a similar setup.

\subsection*{Method and assessment of hydropower dispatch model}\label{sec:hydro_inflow}

For each time step, the state of the remaining energy system is assessed through the net load calculated above. If the net load is negative, i.e., there is no stress on the renewable energy system, no storage is used, and the inflow is added to the reservoir storage. The only exception to this rule is if the storage is full, in which case the inflow is used instantaneously. However, if the net load is positive, indicating a stressful time step, the stored hydropower can be used to lower the net load, under the following constraints:

\begin{itemize}
    \item If the reservoir levels are higher than the mean plus one standard deviation of that calendar day, the hydropower reservoir can operate at maximum capacity
    \item If the reservoir levels are within the mean $\pm$ one standard deviation of that calendar day, the hydropower reservoir can operate at 50\% capacity
    \item If the reservoir levels are lower than the mean minus one standard deviation of that calendar day, the hydropower reservoir can operate at 10\% capacity
\end{itemize}

This gradual decrease of operation capacity is set to avoid a full depletion of the storage when the net load is negative over longer periods of time.

To calculate the mean and standard deviation of the storage per time step and country, we let an energy system optimization framework, ZEN-garden \citep{mannhardt_zen-garden_2025}, provide optimized storage levels for each of the 60 years of each climate period. These storage levels, as well as installed capacities of other generation technologies, are calculated based on the entire converted climate model data set, the process of which is detailed in \citet{de_marco_climate-resilient_2025}. Additionally, the maximum storage capacity is set as the maximum storage level found over all time steps, while the maximum operation capacity for each country was taken from the installed capacities in 2024 as reported by ENTSO-e \citep{entso-e_entso-e_2019}. 

\begin{figure}[h]
    \includegraphics[width=\textwidth]{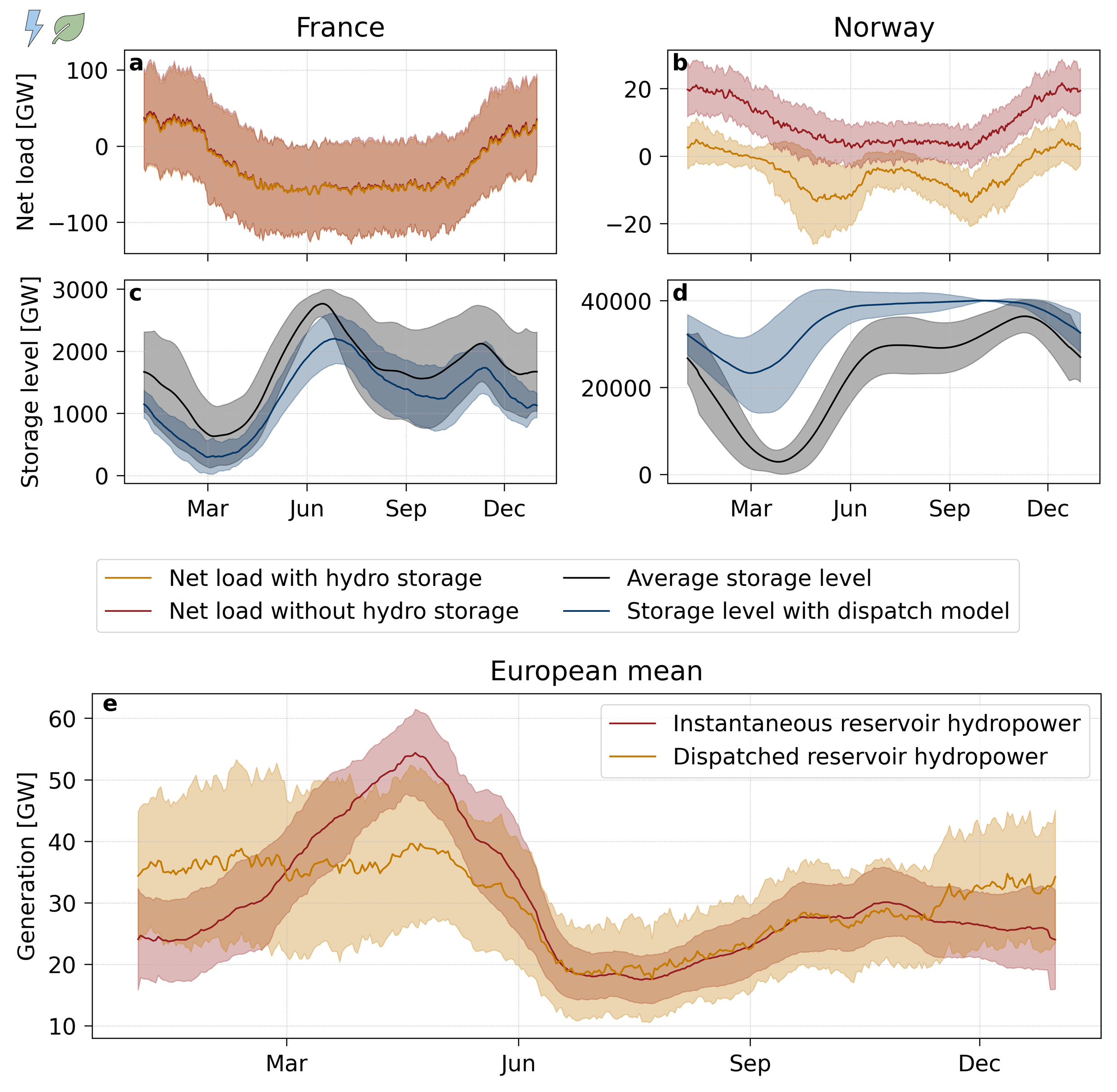}
    \caption{\textbf{Seasonal cycle of (\textbf{a,b}) net load, (\textbf{c,d}) storage levels and (\textbf{e}) changes in the European net load due to reservoir hydropower, for the historical climate period under the scenario "net-zero system; electric heating" (highlighted by upper-left side icons)}. Net load with and without the hydropower dispatch model is shown for (\textbf{a}) France and (\textbf{b}) Norway. Storage levels of the hydropower reservoir is compared to the average storage levels for (\textbf{c}) France and (\textbf{d}) Norway. The shaded areas represent $\pm$ 1 standard deviation. }
    \label{fig:hydro_storage}
\end{figure} 

Figure \ref{fig:hydro_storage} shows the effect of the dispatch model for reservoir hydropower, for the historical climate period under the scenario "net-zero system; electric heating". First, the mean $\pm$ standard deviation of net load and storage levels are shown for two example countries: In panel \textbf{a,c}, France represents the profile of a country where reservoir hydropower plays a minor role in the total net load. While the hydropower storage very slightly reduces the mean net load in the winter season (when net load is, on average, above 0), the overall seasonal cycle of net load is very similar with and without reservoir hydropower. The storage levels in panel \textbf{c} do, however, follow the average storage levels well, with the mean storage level never reaching 0 nor the maximum capacity level. 

In panel \textbf{b,d} Norway represents the profile of a country where reservoir hydropower is one of the major generation technologies. Net load with and without reservoir hydropower show substantial differences, with the hydropower removing most of the winter peaks, but also reducing load in summer, indicating full storage levels and instantaneous inflow usage. This can also be seen in panel \textbf{d}, where storage levels flatten out over the summer, showing storage levels at capacity.

Finally, the overall changes in net load over all the European countries due to the dispatch model can be seen in Figure \ref{fig:hydro_storage}\textbf{e}. The red line shows the average difference in net load if the reservoir hydropower were used instantaneously. Here, we see a clear seasonal cycle, corresponding to higher inflow (and thus lower net load) period during the snow melt season in spring, minimal inflow in late summer, and slight inflow increases in fall and winter. The yellow line, showing the difference in net load when using the dispatch model, diverges from this seasonal cycle in winter, fall and spring --- the increase in inflow in spring is stored, and used in periods of higher net load in fall and winter. 

In summary, this simple, yet powerful dispatch model can simulate hydropower storage effects to a satisfying degree, decreasing peaks in net load during periods of low reservoir inflow.

\subsection*{Method and assessment of transmission effects between countries}\label{sec:transm_app}

We here develop a simple transmission model based on \citet{wohland_more_2017}: Energy can be transmitted between countries, up to the technical limits of the existing Net Transfer Capacities (NTC) \citep{entso-e_entso-e_2019}. One dispatch optimization is run independently for each time step. Countries with negative net load (extra production) can export and/or curtail the excess energy. Countries with positive net load (production is not enough to supply all demand) can import energy from neighbors; the remaining demand not supplied must be produced by backup generators. The objective is to minimize backup generation over all nodes. This results in an optimization problem where transmission flows are limited by NTC constraints, each country must satisfy its nodal energy balance, and backup generation is used only as a last resort. We use the optimal objective, i.e., total backup generation, as the new European net load, after re-dispatch.

\begin{figure}[h]
    \includegraphics[width=\textwidth]{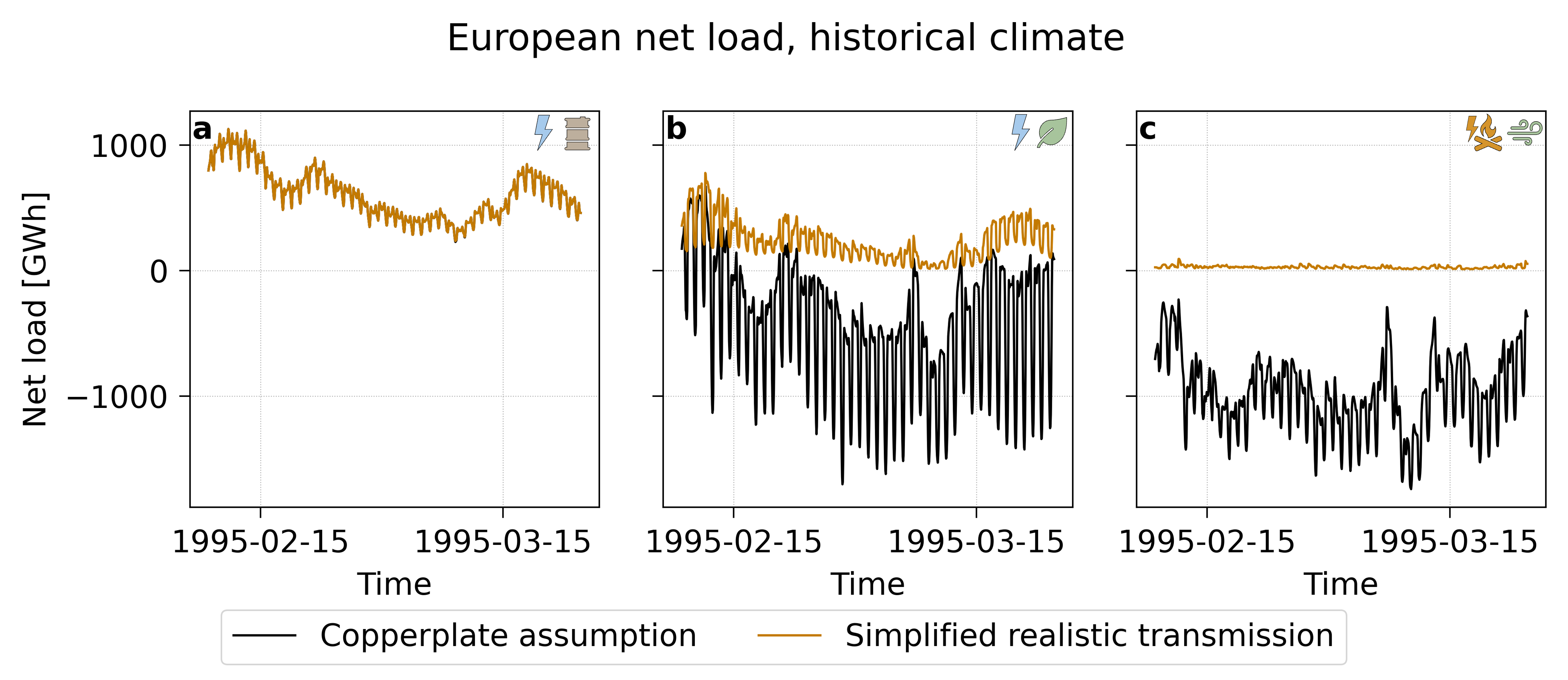}
    \caption{\textbf{Difference in European net load between a copperplate assumption (black line) and simplified realistic transmission effects (yellow line), for different scenarios during the winter-spring transitional period in a historical climate.} scenarios are highlighted by icons in the upper-right corner, and correspond to (\textbf{a}) "historical system; electric heating", (\textbf{b}) "net-zero capacity; electric heating" and (\textbf{c}) "net-zero system, high wind; mixed heating".}
    \label{fig:transmission}
\end{figure} 

Figure \ref{fig:transmission} shows net load for the copperplate assumption and simplified realistic transmission model for three representative scenarios over the transitional period between winter and spring of the model year 1995 in the historical climate.

Since the different scenarios lead to different values of net load, the effect of limited transmission varies substantially: in panel \textbf{a}, net load for the scenario "historical system; electric heating demand" is shown. Here, due to the lower installed capacity of renewables, and the importance of cold season heating demand, net load is consistently above zero during this period. This means that the simplified realistic transmission between countries has no visible effect on the net load, since there is little to no negative net load for countries to transmit. 

On the other hand, panel \textbf{c} shows net load for the scenario "high wind, net-zero system; mixed heating demand", which is consistently below zero during this period. This can be explained in an opposite manner to panel \textbf{a}, since heating demand is less important here, and the net-zero system setups have higher installed capacities of renewables. This panel shows that when net load is below zero, the transmission essentially flattens the time series to zero, except for the time steps where at least one country presents a net load above zero that cannot be compensated for by the neighboring countries.

Finally, panel \textbf{b} shows the scenario "net-zero system; electric heating demand", where net load has peaks and troughs above and below zero. Here, we see that certain net load peaks, like those around 15 February, are not substantially exacerbated when adding transmission effects. The peaks after 15 March, however, become substantially more intense, with increases not being proportional to the copperplate peak values.

This reveals non-linear mechanisms of transmission that can increase the intensity of positive net load, depending on the each country's contribution to said net load. However, there is a large variance of net load values across scenarios, with certain scenarios showing prolonged periods below zero. Such scenarios likely underestimate certain sources of demand, and the simplified realistic transmission model thus removes most of the temporal variation in net load.

\subsection*{Appendix result figures}

\begin{figure}[h]
 \centering
        \includegraphics[width=\textwidth]{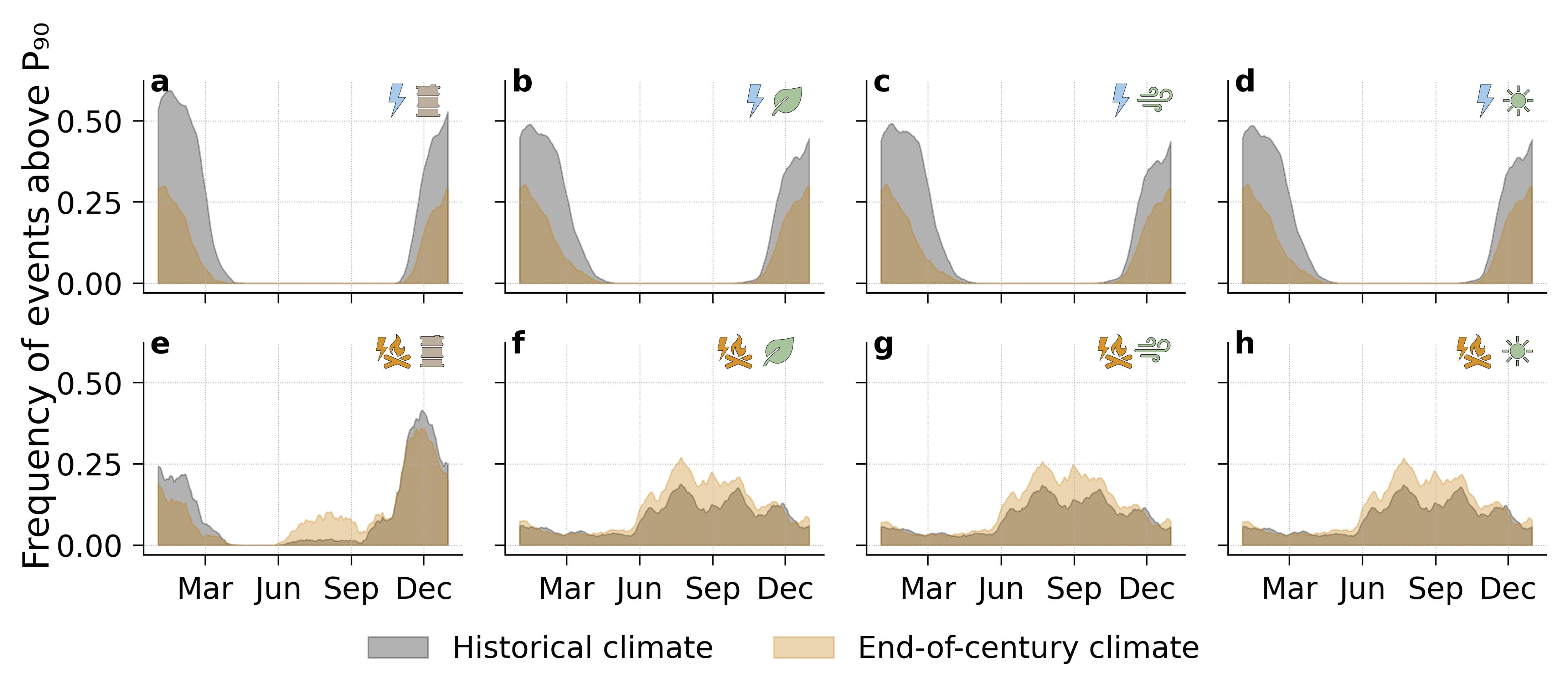}
 \caption{\textbf{Seasonal cycle of high net load events (above $\mathrm{P}_{90}$ ) for different energy scenarios, with simplified realistic transmission.} Results for historical climate conditions are shown in black and results using end-of-century climate are shown in orange. Subpanels are organized as in Fig. 1, icons in the upper-right corner representing energy scenario choice. Specifically, they correspond to the (\textbf{a-d}) electric heating and (\textbf{e-h}) mixed heating setups, and the (\textbf{a,e}) historical system, (\textbf{b,f}) net-zero TYNDP system, (\textbf{c,g}) net-zero, high wind system and (\textbf{d,h}) net-zero, high solar system setups.
 }
\label{fig:season_peaks}
\end{figure}

\begin{figure}[h]
 \centering
        \includegraphics[width=\textwidth]{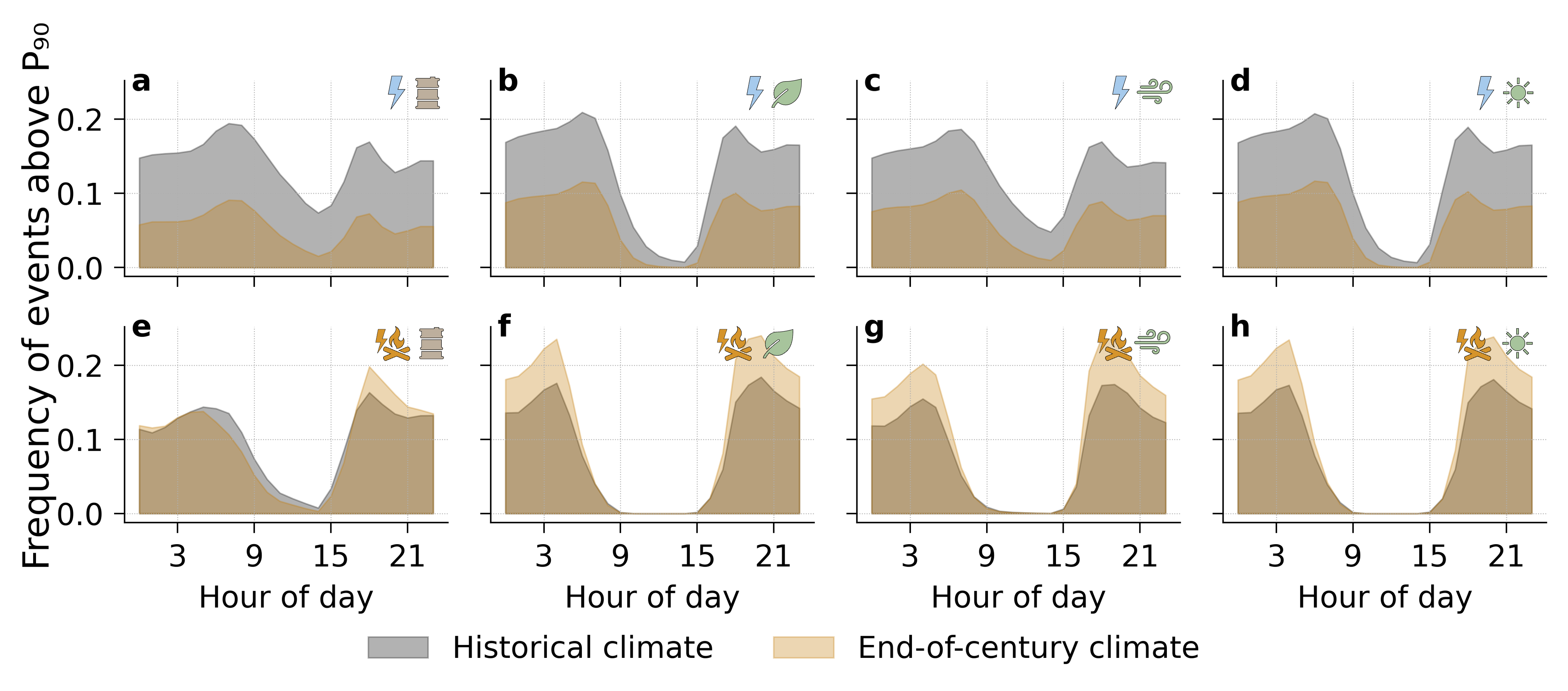}
 \caption{\textbf{Diurnal cycle of high net load events (above $\mathrm{P}_{90}$ ) for different energy scenarios, with simplified realistic transmission.} Results for historical climate conditions are shown in black and results using end-of-century climate are shown in orange. Subpanels are organized as in Fig. 1, icons in the upper-right corner representing energy scenario choice. Specifically, they correspond to the (\textbf{a-d}) electric heating and (\textbf{e-h}) mixed heating setups, and the (\textbf{a,e}) historical system, (\textbf{b,f}) net-zero TYNDP system, (\textbf{c,g}) net-zero, high wind system and (\textbf{d,h}) net-zero, high solar system setups.
 }
\label{fig:daily_peaks}
\end{figure}

\begin{figure}[h]
 \centering
        \includegraphics[width=\textwidth]{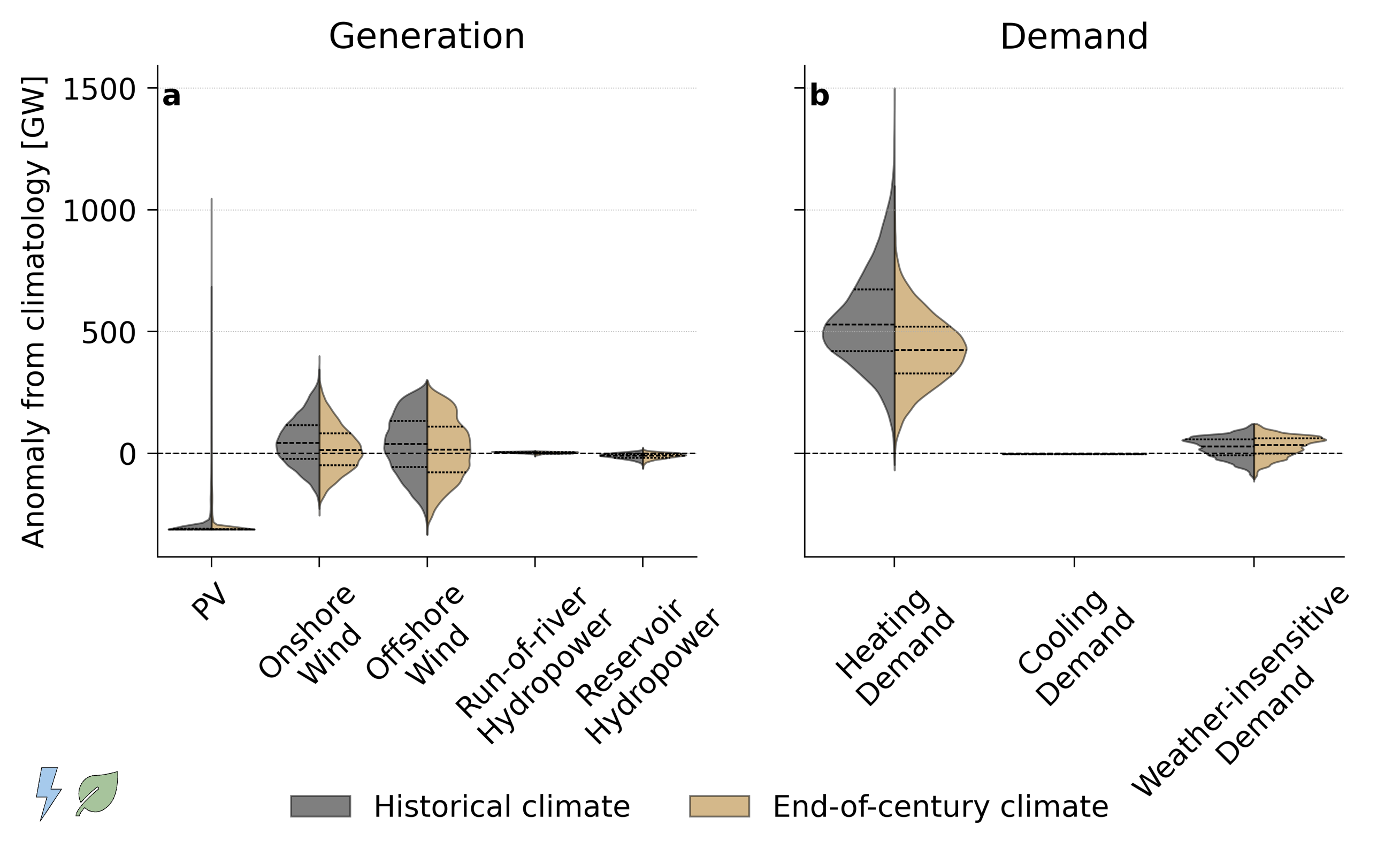}
 \caption{\textbf{Violin plots of (a) generation and (b) demand anomaly for high net load events (above $\mathrm{P}_{90}$), for all seasons.} Results for historical climate conditions are shown in black and results using end-of-century climate are shown in orange. Anomalies are calculated over all available time steps in both climate periods. The scenario corresponds to "Electric heating; net-zero system" (represented by the bottom left icons), using simplified realistic transmission. A negative generation and positive demand technology anomaly means that it contributed to the stress of the event.
 }
\label{fig:violins_net_full}
\end{figure}

\begin{figure}[h]
 \centering
        \includegraphics[width=\textwidth]{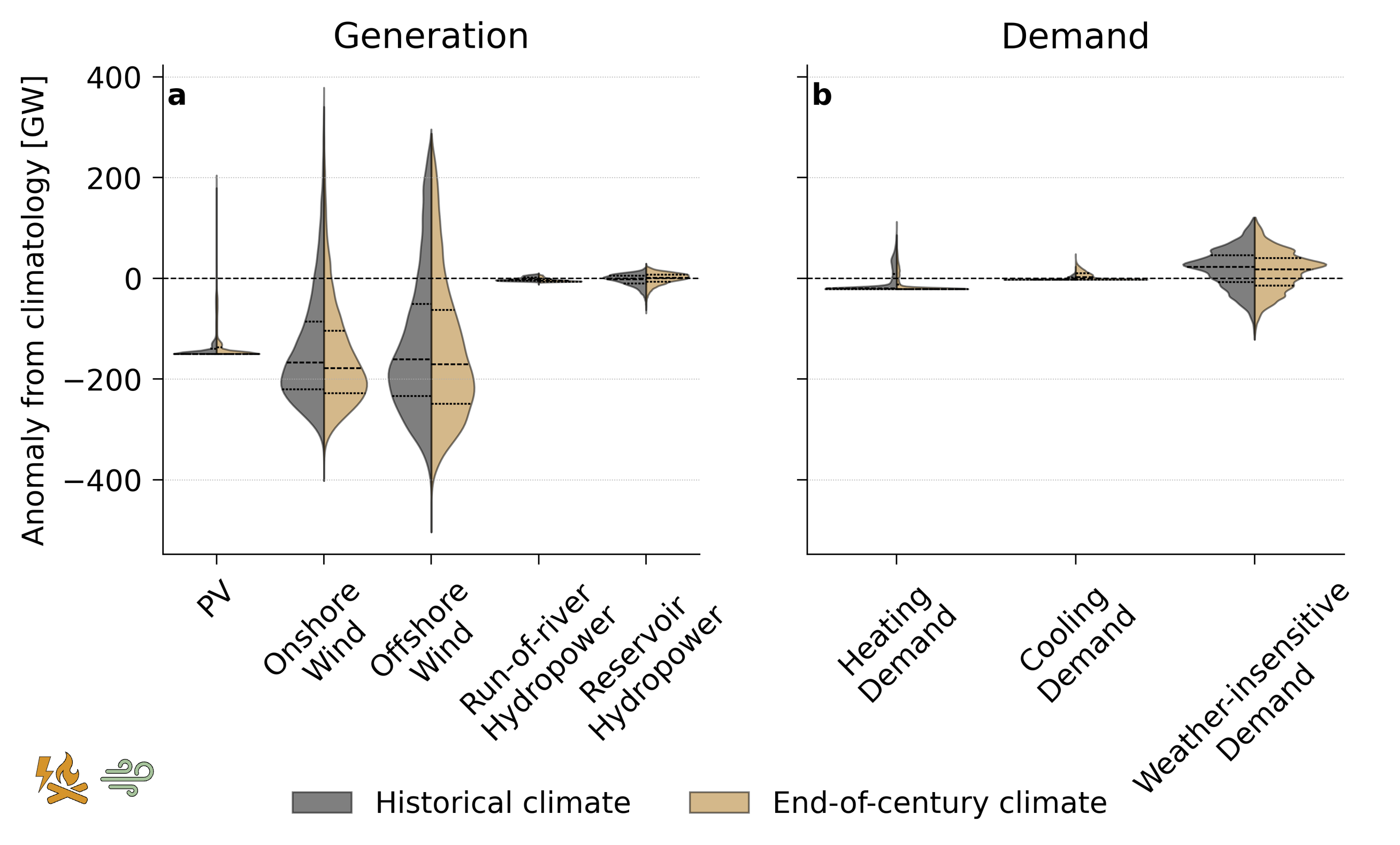}
 \caption{\textbf{Violin plots of (a) generation and (b) demand anomaly for high net load events (above $\mathrm{P}_{90}$), for all seasons.} Results for historical climate conditions are shown in black and results using end-of-century climate are shown in orange. Anomalies are calculated over all available time steps in both climate periods. The scenario corresponds to "Mixed heating; net-zero, high wind system" (represented by the bottom left icons), using simplified realistic transmission. A negative generation and positive demand technology anomaly means that it contributed to the stress of the event.
 }
\label{fig:violins_wind_curr}
\end{figure}


\end{document}